\documentclass[12pt,english]{article}
\usepackage[T1]{fontenc}
\usepackage[latin9]{inputenc}
\usepackage{longtable}
\usepackage{float}
\usepackage{booktabs}
\usepackage{graphicx}
\usepackage{setspace}
\makeatletter

\providecommand{\tabularnewline}{\\}

\usepackage{aStyle}

\makeatother

\usepackage{babel}
\begin{document}
\begin{doublespace}
\begin{center}
\textbf{\Large{}Yield Stages of Viscoplastic Fluids in Tubes of Elliptical,
Rectangular, Triangular and Annular Cross Sections}\vspace{-1.3cm}
\par\end{center}
\end{doublespace}

\begin{center}
Taha Sochi\footnote{University College London - Department of Physics \& Astronomy - Gower
Street - London - WC1E 6BT. Email: t.sochi@ucl.ac.uk.}\vspace{-0.0cm}
\par\end{center}

\noindent \phantomsection \addcontentsline{toc}{section}{Abstract}

\noindent \textbf{Abstract}: In this paper we continue our previous
investigation about the use of stress function in the flow of generalized
Newtonian fluids through conduits of circular and non-circular (or/and
multiply connected) cross sections where we visualize the stages of
yield in the process of flow of viscoplastic fluids through tubes
of elliptical, rectangular, triangular and annular cross sections.
The purpose of this qualitative investigation is to provide an initial
idea about the expected yield development in the process of flow of
yield-stress fluids through tubes of some of the most common non-circular
(and non-simply-connected) cross sectional geometries.\footnote{All symbols are defined in $\S$ \hyperref[Nomenclature]{Nomenclature}
in the back of this paper.}\vspace{0.3cm}


\clearpage{}

\phantomsection \addcontentsline{toc}{section}{Table of Contents}\tableofcontents{}

\clearpage{}

\section{Introduction}

Yield-stress is a complex phenomenon which is difficult to understand,
quantify and model. There are many investigations of various aspects
of the rheology and flow of yield-stress fluids in bulk and \textit{in
situ} within various conditions, contexts, topics, applications and
so on (see for instance \cite{MerrillCP1969,AlfarissP1984,MorrisRSGSB1989,ChaplainMGC1992,Nguyen1992,LiddellB1996,LindnerCB2000,PicardABL2002,ChaseD2003,BalhoffT2004,ChenRY2005,HarteCC2007,CoussotTLO2009,JossicM2009,AlexandrouCG2009,DivouxTBM2010,KaoullasG2013,ChevalierCCDCe2013,FarayolaOA2013,ShahsavariM2016,SalehiRS2019,GargBSHJ2021,PourzahediF2024}).
We also investigated in the past a number of issues about yield stress
fluids (such as their yield condition and the flow rate) in circular
tubes, thin slits and networks of circular tubes (see for instance
\cite{SochiThesis2007,SochiB2008,SochiYield2010,SochiElasticYield2013,SochiYieldBal2013})
as part of our interest in fluid mechanics in general and non-Newtonian
fluids in particular (especially the flow of generalized Newtonian
fluids in tubes, slits and networks of interconnected tubes).

However, we are not aware of systematic investigations of the stages
of yield of yield-stress fluids in tubes of non-circular (or multiply
connected) tubes. In this study we employ the idea of stress function
which we propose to use previously (see \cite{SochiStress1D2015,SochiStressFunc22015})
to investigate the flow of generalized Newtonian fluids in conduits
of various cross sectional geometries (i.e., circular, non-circular
with simple or multiple connectivity). The study is simply based on
visualizing the stress function (and hence the development of yield
stages which is based on the stress function) of tubes of four cross
sectional geometries: ellipse, rectangle, equilateral triangle, and
circular annulus.

We recognize that yield stress is a very complex phenomenon (see for
instance \cite{BarnesW1985,Astarita1990,Evans1992,Barnes1999,MollerMB2006,Renardy2010})
and hence this study should provide no more than a rough initial qualitative
idea about the stages of yield. However, it should be very useful
for those who investigate this subject since these visualizations
provide them with some rough ideas and rules of thumb that may be
used to start their investigation and monitor the progress since such
ideas and rules can prevent making gross errors and bad judgments.

\clearpage{}

\section{Ellipse}

\begin{figure}[H]
\centering\includegraphics[scale=0.9]{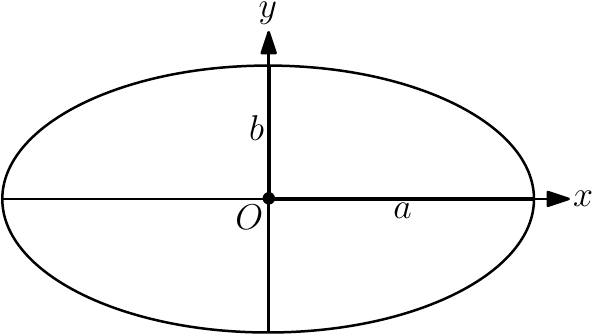}\label{FigEllipse}\caption{The setting of the elliptical cross section of the tube where $a$
and $b$ represent the semi-major and semi-minor axes and $O$ is
the origin of coordinates. The $z$ axis is emanating from the origin
and is perpendicular to the plane of cross section.}
\end{figure}

\begin{figure}
\centering\includegraphics[scale=0.9]{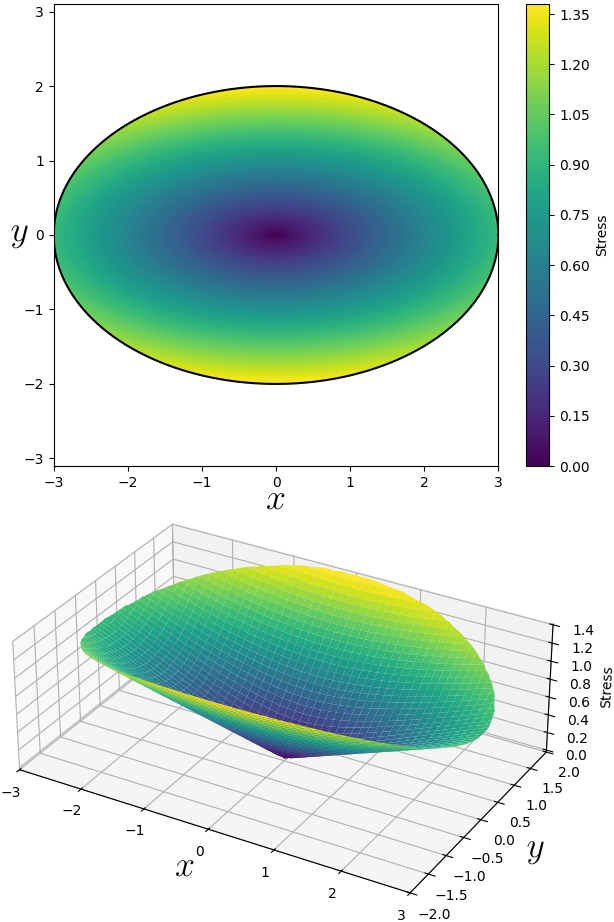}\label{FigStressEllipse}\caption{2D and 3D visualizations of the stress function for an elliptical
tube with $a=3$ and $b=2$ with a typical pressure gradient.}
\end{figure}

\begin{figure}
\centering\includegraphics[scale=0.327]{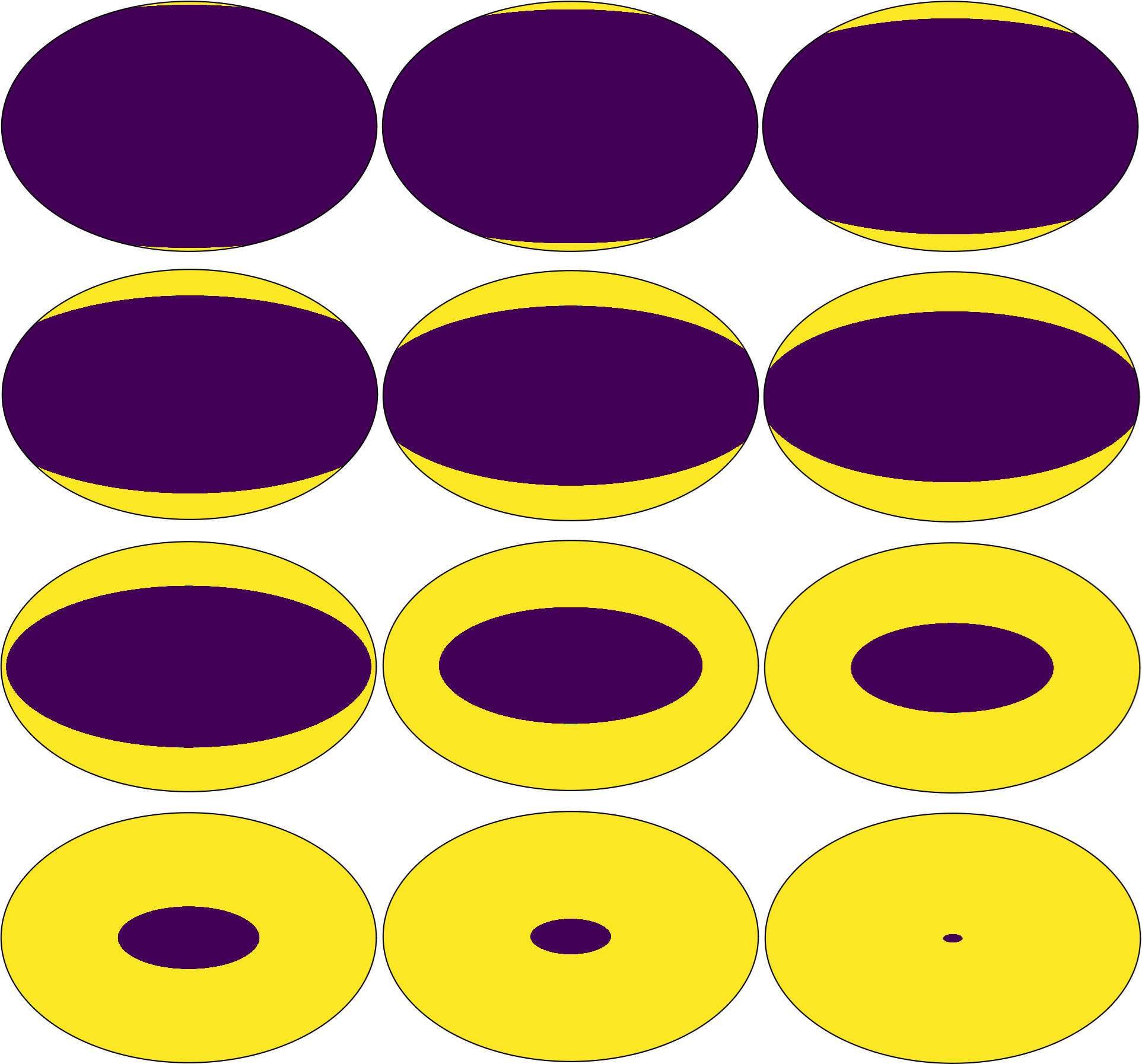}\label{FigStagesEllips}\caption{Visualization of the stages of yield for a tube of elliptical cross
section.}
\end{figure}

For a conduit centered on the origin of coordinates with an elliptical
cross section of semi-major axis $a$ along the $x$ axis and semi-minor
axis $b$ along the $y$ axis (refer to Figure \ref{FigEllipse})
the components of the stress function are given by:

\begin{equation}
\tau_{xz}=-\frac{\partial p}{\partial z}\frac{b^{2}x}{a^{2}+b^{2}}
\end{equation}

\begin{equation}
\tau_{yz}=-\frac{\partial p}{\partial z}\frac{a^{2}y}{a^{2}+b^{2}}
\end{equation}
The magnitude of this stress function is visualized in Figure \ref{FigStressEllipse}
and the yield stages are visualized in Figure \ref{FigStagesEllips}.\vspace{5cm}

\noindent \clearpage{}

\noindent \clearpage{}

\noindent \clearpage{}

\section{Rectangle}

\begin{figure}[H]
\centering\includegraphics[scale=0.9]{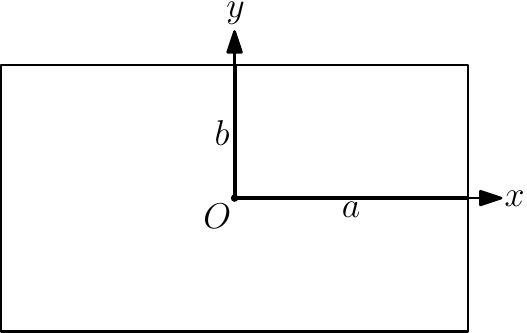}\label{FigRectangle}\caption{The setting of the rectangular cross section of the tube where $a$
and $b$ represent the semi-length and semi-width and $O$ is the
origin of coordinates. The $z$ axis is emanating from the origin
and is perpendicular to the plane of cross section.}
\end{figure}

\begin{figure}
\centering\includegraphics[scale=0.74]{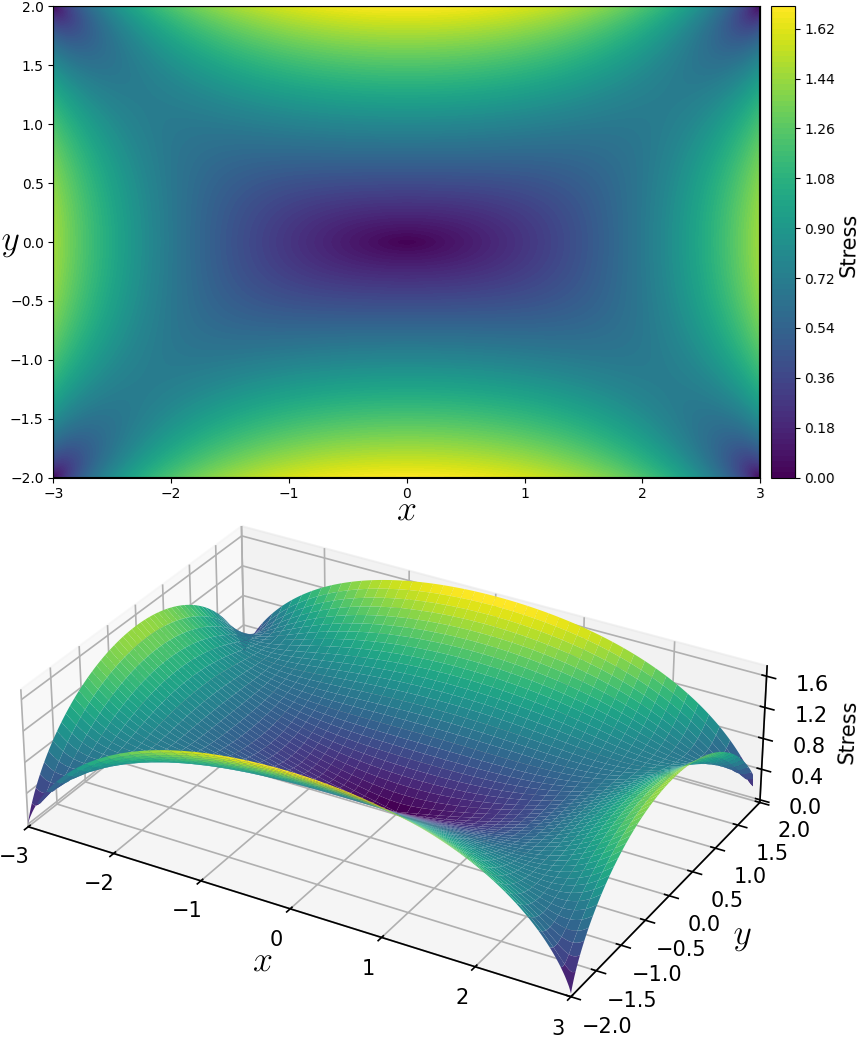}\label{FigStressRect}\caption{2D and 3D visualizations of the stress function for a rectangular
tube with $a=3$ and $b=2$ with a typical pressure gradient.}
\end{figure}

\begin{figure}
\centering\includegraphics[scale=0.68]{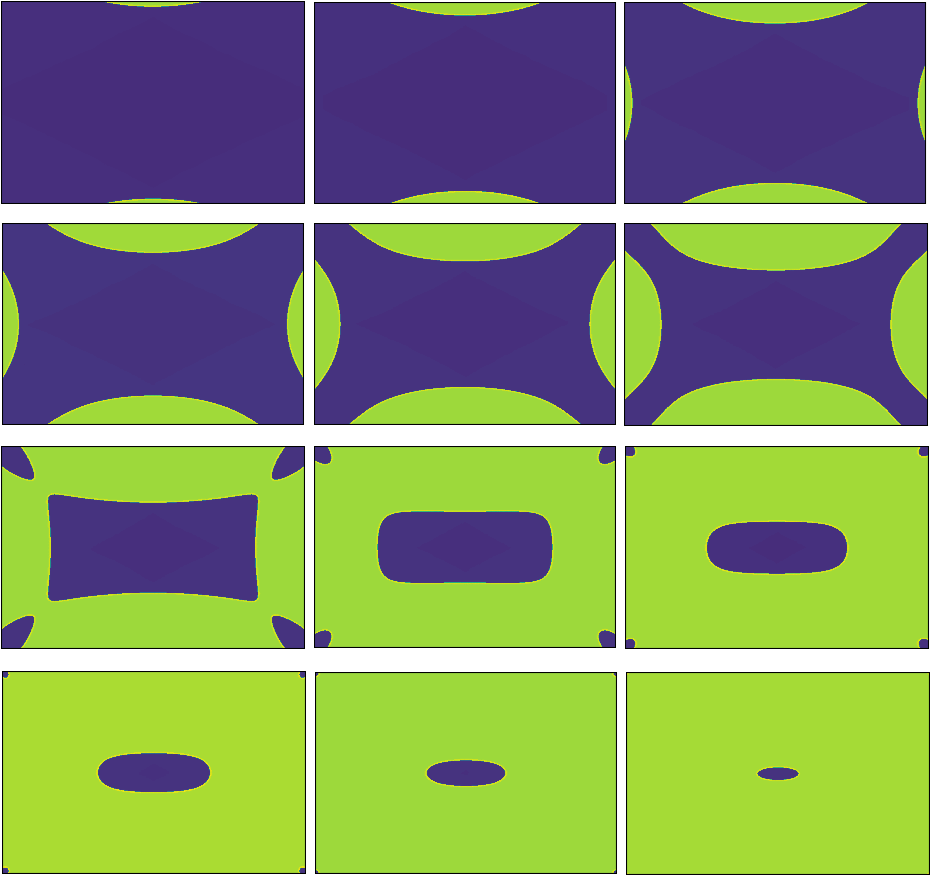}\label{FigStagesRect}\caption{Visualization of the stages of yield for a tube of rectangular cross
section.}
\end{figure}

For a conduit centered on the origin of coordinates with a rectangular
cross section of half length $a$ along the $x$ axis and half width
$b$ along the $y$ axis (refer to Figure \ref{FigRectangle}) the
components of the stress function are given by:

\begin{equation}
\tau_{xz}=-\frac{\partial p}{\partial z}\frac{8b}{\pi^{2}}\sum_{i=1,3,5,\ldots}^{\infty}\frac{\left(-1\right)^{\left(i-1\right)/2}}{i^{2}}\frac{\sinh\left(i\pi x/2b\right)}{\cosh\left(i\pi a/2b\right)}\cos\left(i\pi y/2b\right)
\end{equation}

\begin{equation}
\tau_{yz}=-\frac{\partial p}{\partial z}\left[y-\frac{8b}{\pi^{2}}\sum_{i=1,3,5,\ldots}^{\infty}\frac{\left(-1\right)^{\left(i-1\right)/2}}{i^{2}}\frac{\cosh\left(i\pi x/2b\right)}{\cosh\left(i\pi a/2b\right)}\sin\left(i\pi y/2b\right)\right]
\end{equation}
The magnitude of this stress function is visualized in Figure \ref{FigStressRect}
and the yield stages are visualized in Figure \ref{FigStagesRect}.\vspace{5cm}

\noindent \clearpage{}

\noindent \clearpage{}

\clearpage{}

\section{Triangle}

\begin{figure}[H]
\centering\includegraphics[scale=0.9]{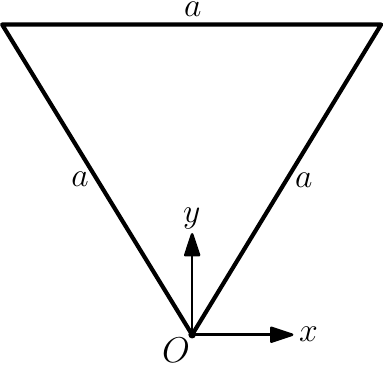}\label{FigTriangle}\caption{The setting of the triangular cross section of the tube where $a$
represents the length of each side of the triangle and $O$ is the
origin of coordinates. The $z$ axis is emanating from the origin
and is perpendicular to the plane of cross section.}
\end{figure}

\begin{figure}
\centering\includegraphics[scale=0.85]{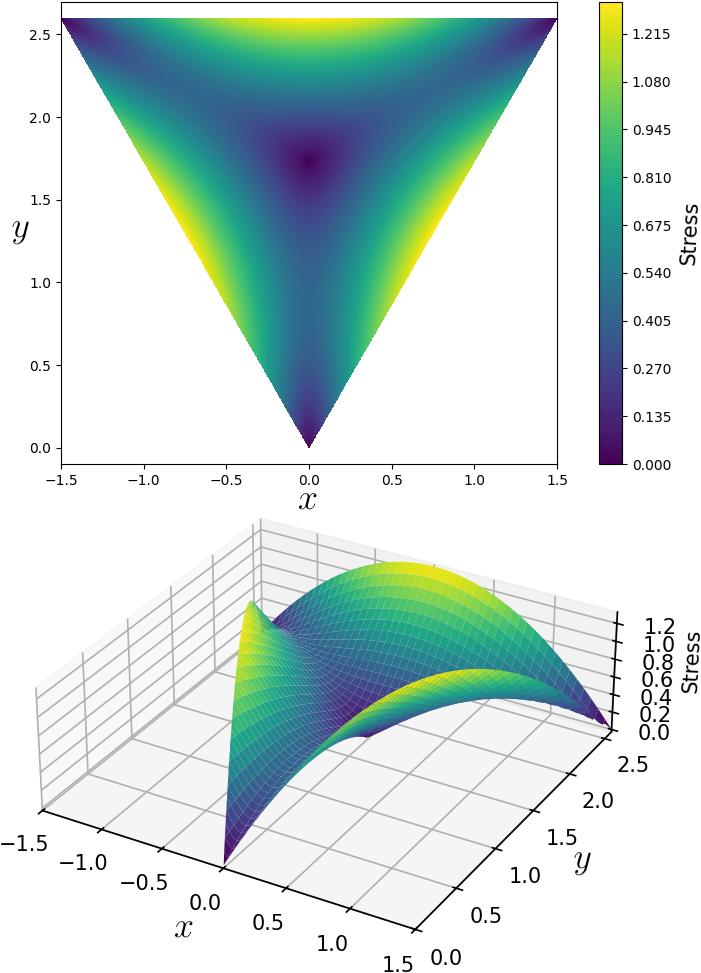}\label{FigStressTriang}\caption{2D and 3D visualizations of the stress function for an equilateral
triangular tube with $a=3$ with a typical pressure gradient.}
\end{figure}

\begin{figure}
\centering\includegraphics[scale=0.92]{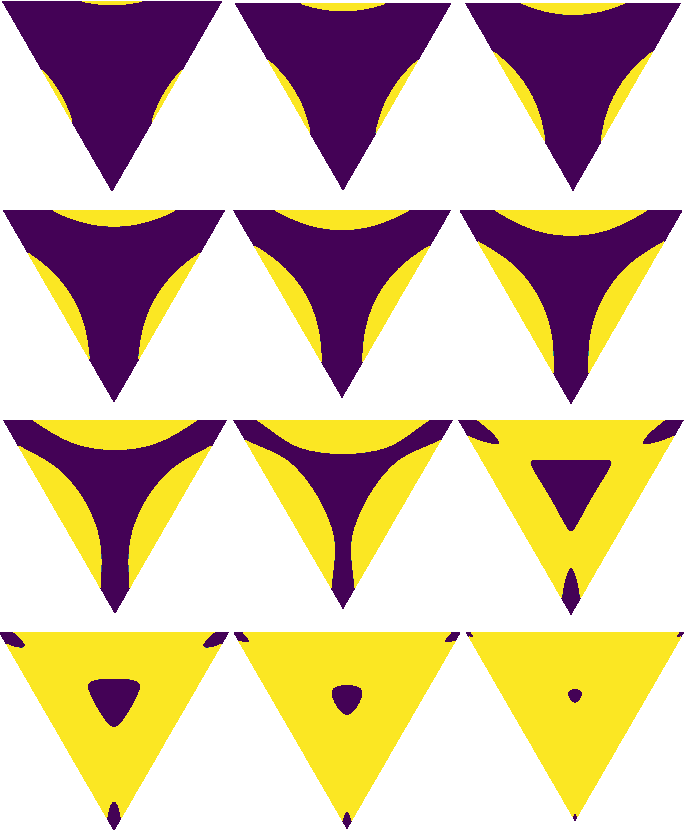}\label{FigStagesTriang}\caption{Visualization of the stages of yield for a tube of triangular cross
section.}
\end{figure}

For a conduit with an equilateral triangular cross section of side
$a$ in the coordinates system given in Figure \ref{FigTriangle}
the components of the stress function are given by:

\begin{equation}
\tau_{xz}=-\frac{\partial p}{\partial z}\frac{\sqrt{3}}{a}\left(\frac{a\sqrt{3}}{2}-y\right)x
\end{equation}

\begin{equation}
\tau_{yz}=-\frac{\partial p}{\partial z}\frac{1}{2\sqrt{3}a}\left(-3x^{2}+3y^{2}-a\sqrt{3}y\right)
\end{equation}
The magnitude of this stress function is visualized in Figure \ref{FigStressTriang}
and the yield stages are visualized in Figure \ref{FigStagesTriang}.\vspace{5cm}

\noindent \clearpage{}

\noindent \clearpage{}

\clearpage{}

\section{Annulus}

\begin{figure}[H]
\centering\includegraphics[scale=0.9]{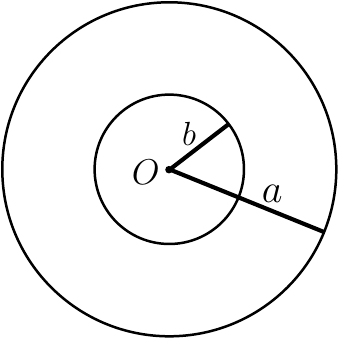}\label{FigAnnulus}\caption{The setting of the annular cross section of the tube where $a$ and
$b$ represent the outer and inner radii of the annulus and $O$ is
the origin of coordinates. The $z$ axis is emanating from the origin
and is perpendicular to the plane of cross section.}
\end{figure}

\begin{figure}
\centering\includegraphics[scale=0.75]{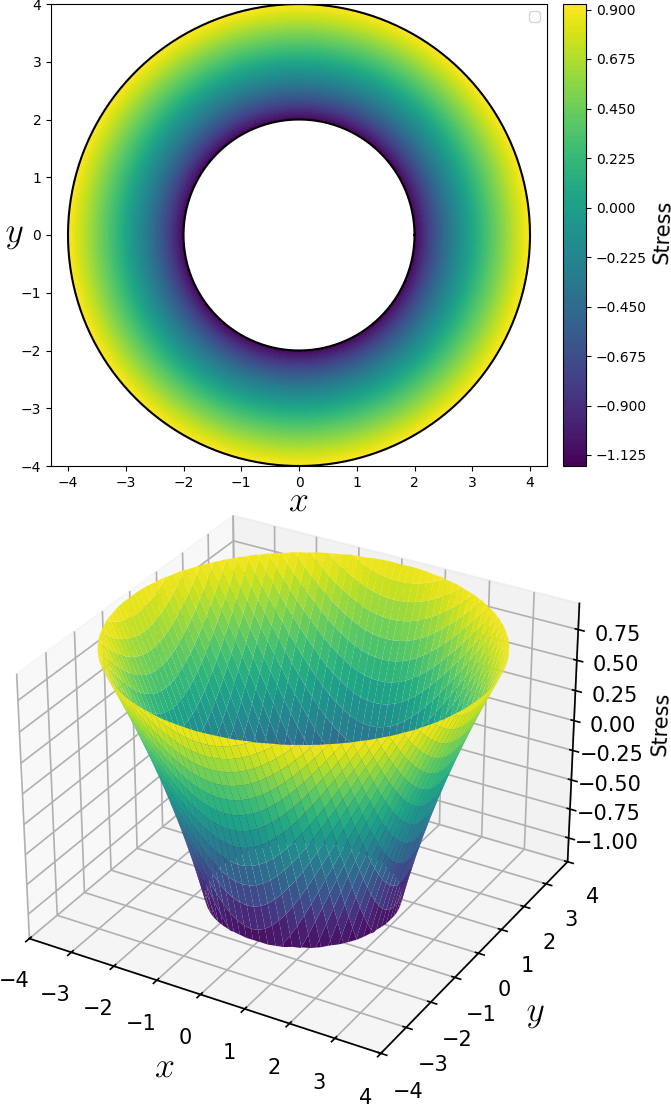}\label{FigStressAnn}\caption{2D and 3D visualizations of the stress function for an annular tube
with $a=4$ and $b=2$ with a typical pressure gradient.}
\end{figure}

\begin{figure}
\centering\includegraphics[scale=0.75]{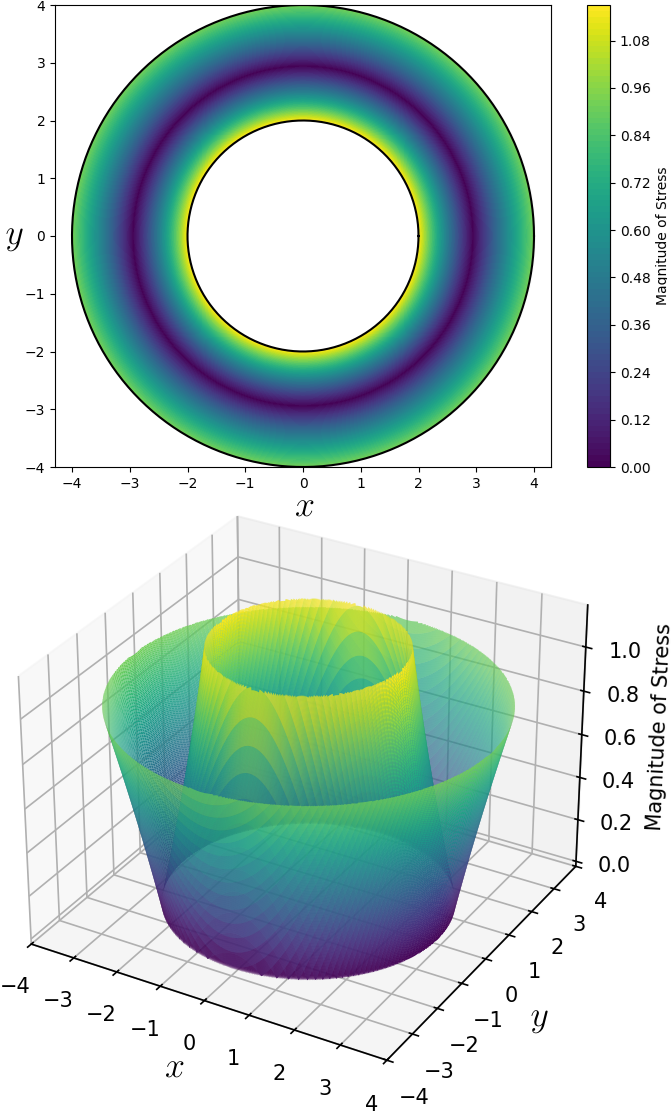}\label{FigStressAnnMag}\caption{2D and 3D visualizations of the magnitude of the stress function for
an annular tube with $a=4$ and $b=2$ with a typical pressure gradient.}
\end{figure}

\begin{figure}
\centering\includegraphics[scale=0.415]{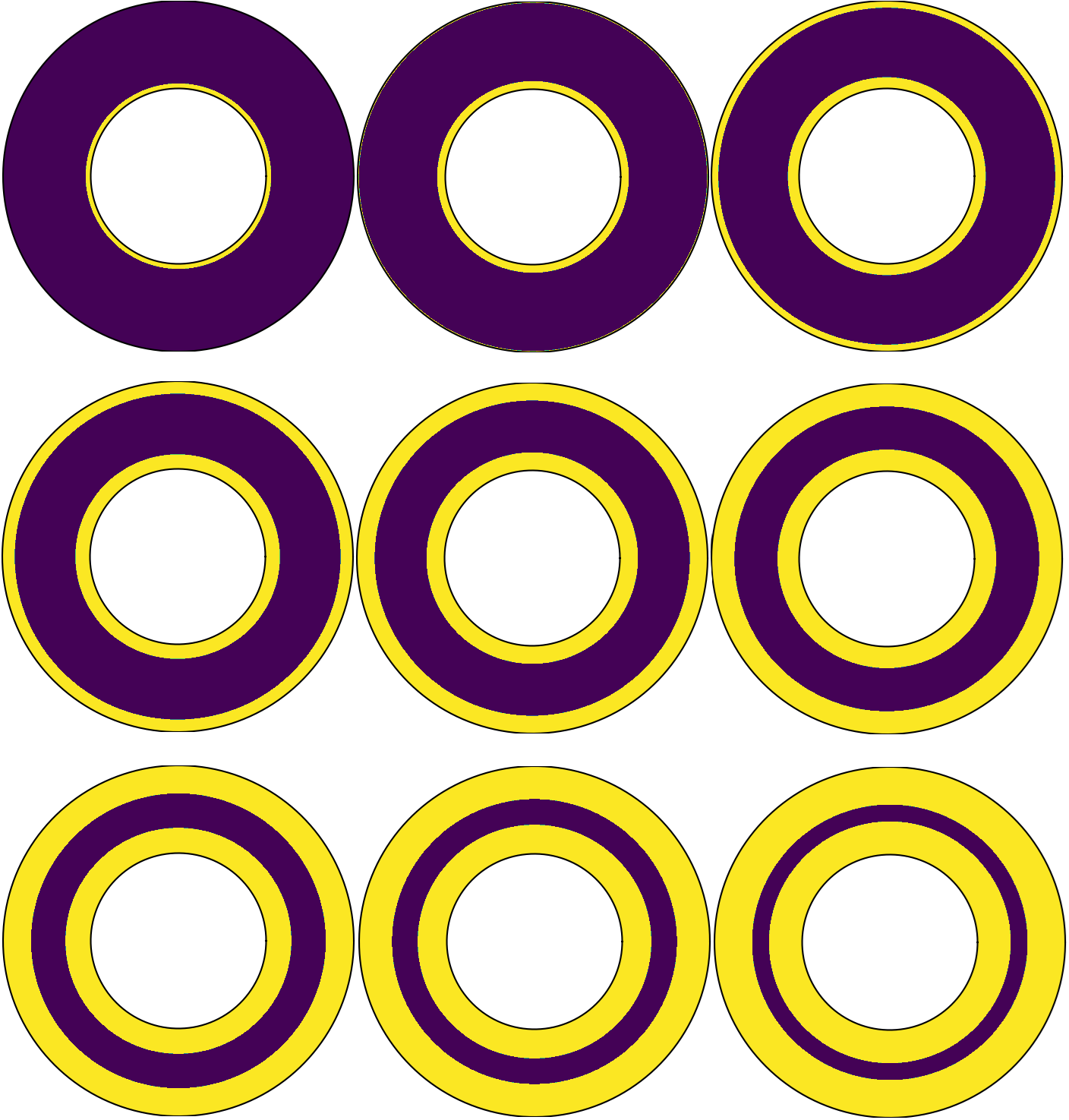}\label{FigStagesAnn}\caption{Visualization of the stages of yield for a tube of annular cross section.}
\end{figure}

For a concentric circular annulus with an outer radius $a$ and an
inner radius $b$ (refer to Figure \ref{FigAnnulus}), using a cylindrical
coordinates system whose $z$-axis is oriented along the annulus axis
of symmetry, the components of the stress function are given by:

\begin{equation}
\tau_{rz}=-\frac{\partial p}{\partial z}\frac{1}{4}\left[2r+\frac{\left(a^{2}-b^{2}\right)}{\ln\left(b/a\right)}\frac{1}{r}\right]
\end{equation}

\begin{equation}
\tau_{\theta z}=0
\end{equation}
This stress function is visualized in Figure \ref{FigStressAnn}.\footnote{Actually, what is visualized is $\tau_{rz}$ with sign reversal.}
An interesting thing about the annulus stress function is that it
has opposite signs in the regions at the proximity of the two walls
(i.e. inner and outer walls). However, a logical assumption is that
the yield depends on the magnitude of the shear stress where this
magnitude is visualized in Figure \ref{FigStressAnnMag}. So, according
to this ``stress function'' (which represents the magnitude) the
yield stages will look like what is visualized in Figure \ref{FigStagesAnn}.\vspace{2.5cm}

\noindent \clearpage{}

\noindent \clearpage{}

\noindent \clearpage{}

\clearpage{}

\section{\label{secConclusions}Conclusions}

We outline in the following points the main achievements and conclusions
of the present paper:
\begin{enumerate}
\item We continued our previous investigations about yield stress phenomenon
in the flow of conduits and networks by inspecting the expected yield
development for the flow of yield-stress fluids in tubes of non-circular
or multiply connected cross sections in a number of geometries (i.e.
ellipse, rectangle, equilateral triangle, and circular annulus) by
using the idea of stress function.
\item We visualized the stress function and the stages of yield for the
flow of yield-stress fluids through tubes of the aforementioned four
cross sectional geometries, i.e. elliptical, rectangular, equilateral
triangular, and circular annular. Inspection of these visualizations
indicates their sensibility and usefulness.
\item This investigation should provide a general idea about the yield condition
and the yield stages and hence it can be used as a starter for more
complex investigations of quantitative nature where velocity profile
and volumetric flow rate (as well as other physical quantities and
attributes) can be considered.
\item We recognize that yield-stress phenomenon is more complex to be investigated
and analyzed properly and rigorously by this simplistic visualization
method. However, our investigation should provide a clear idea about
the nature of the stress function of the investigated geometries.
Moreover, it should provide a reasonable starter for investigating
the yield condition and the stages of yield development in these geometries.
\item We look for more investigations of this type by other researchers
in this field to improve our understanding of the yield-stress phenomenon.
\end{enumerate}
\phantomsection 
\addcontentsline{toc}{section}{References}\bibliographystyle{unsrt}
\bibliography{YieldVisNonCirc}

\begin{thebibliography}{10}

\bibitem{MerrillCP1969}
E.W. Merrill; C.S. Cheng;~G.A. Pelletier.
\newblock {Yield stress of normal human blood as a function of endogenous
  fibrinogen}.
\newblock {\em Journal of Applied Physiology}, 26(1):1--3, 1969.

\bibitem{AlfarissP1984}
T.F. Al-Fariss;~K.L. Pinder.
\newblock {Flow of a shear-thinning liquid with yield stress through porous
  media}.
\newblock {\em SPE 13840}, 1984.

\bibitem{MorrisRSGSB1989}
C.L. Morris; D.L. Rucknagel; R. Shukla; R.A. Gruppo; C.M. Smith; P.~Blackshear
  Jr.
\newblock {Evaluation of the yield stress of normal blood as a function of
  fibrinogen concentration and hematocrit}.
\newblock {\em Microvascular Research}, 37(3):323--338, 1989.

\bibitem{ChaplainMGC1992}
V.~Chaplain; P. Mills; G. Guiffant;~P. Cerasi.
\newblock {Model for the flow of a yield fluid through a porous medium}.
\newblock {\em Journal de Physique II}, 2:2145--2158, 1992.

\bibitem{Nguyen1992}
Q.D. Nguyen;~D.V. Boger.
\newblock {Measuring the Flow Properties of Yield Stress Fluids}.
\newblock {\em Annual Review of Fluid Mechanics}, 24:47--88, 1992.

\bibitem{LiddellB1996}
P.V. Liddell;~D.V. Boger.
\newblock {Yield stress measurements with the vane}.
\newblock {\em Journal of Non-Newtonian Fluid Mechanics}, 63(2-3):235--261,
  1996.

\bibitem{LindnerCB2000}
A.~Lindner; P. Coussot;~D. Bonn.
\newblock {Viscous Fingering in a Yield Stress Fluid}.
\newblock {\em Physical Review Letters}, 85(2):314--317, 2000.

\bibitem{PicardABL2002}
G.~Picard; A. Ajdari; L. Bocquet;~F. Lequeux.
\newblock {Simple model for heterogeneous flows of yield stress fluids}.
\newblock {\em Physical Review E}, 66(5):051501, 2002.

\bibitem{ChaseD2003}
G.G. Chase;~P. Dachavijit.
\newblock {Incompressible cake filtration of a yield stress fluid}.
\newblock {\em Separation Science and Technology}, 38(4):745--766, 2003.

\bibitem{BalhoffT2004}
M.T. Balhoff;~K.E. Thompson.
\newblock {Modeling the steady flow of yield-stress fluids in packed beds}.
\newblock {\em AIChE Journal}, 50(12):3034--3048, 2004.

\bibitem{ChenRY2005}
M.~Chen; W.R. Rossen;~Y.C. Yortsos.
\newblock {The flow and displacement in porous media of fluids with yield
  stress}.
\newblock {\em Chemical Engineering Science}, 60(15):4183--4202, 2005.

\bibitem{HarteCC2007}
F.~Harte; S. Clark;~G.V. Barbosa-C\'{a}novas.
\newblock {Yield stress for initial firmness determination on yogurt}.
\newblock {\em Journal of Food Engineering}, 80(3):990--995, 2007.

\bibitem{CoussotTLO2009}
P.~Coussot; L. Tocquer; C. Lanos;~G. Ovarlez.
\newblock {Macroscopic vs. local rheology of yield stress fluids}.
\newblock {\em Journal of Non-Newtonian Fluid Mechanics}, 158(1-3):85--90,
  2009.

\bibitem{JossicM2009}
L.~Jossic;~A. Magnin.
\newblock {Drag of an isolated cylinder and interactions between two cylinders
  in yield stress fluids}.
\newblock {\em Journal of Non-Newtonian Fluid Mechanics}, 164(1-3):9--16, 2009.

\bibitem{AlexandrouCG2009}
A.N. Alexandrou; N. Constantinou;~G. Georgiou.
\newblock {Shear rejuvenation, aging and shear banding in yield stress fluids}.
\newblock {\em Journal of Non-Newtonian Fluid Mechanics}, 158(1-3):6--17, 2009.

\bibitem{DivouxTBM2010}
T.~Divoux; D. Tamarii; C. Barentin;~S. Manneville.
\newblock {Transient Shear Banding in a Simple Yield Stress Fluid}.
\newblock {\em Physical Review Letters}, 104(20):208301, 2010.

\bibitem{KaoullasG2013}
G.~Kaoullas;~G.C. Georgiou.
\newblock {Newtonian Poiseuille flows with slip and non-zero slip yield
  stress}.
\newblock {\em Journal of Non-Newtonian Fluid Mechanics}, 197:24--30, 2013.

\bibitem{ChevalierCCDCe2013}
T.~Chevalier; C. Chevalier; X. Clain; J.C. Dupla; J. Canou; S. Rodts;~P.
  Coussot.
\newblock {Darcy's law for yield stress fluid flowing through a porous medium}.
\newblock {\em Journal of Non-Newtonian Fluid Mechanics}, 195:57--66, 2013.

\bibitem{FarayolaOA2013}
K.K. Farayola; A.T. Olaoye;~A. Adewuyi.
\newblock {Petroleum Reservoir Characterisation for Fluid with Yield Stress
  Using Finite Element Analyses}.
\newblock {\em Nigeria Annual International Conference and Exhibition, 30 July
  - 1 August 2013, Lagos, Nigeria}, 2013.

\bibitem{ShahsavariM2016}
S.~Shahsavari;~G.H. McKinley.
\newblock {Mobility and pore-scale fluid dynamics of rate-dependent
  yield-stress fluids flowing through fibrous porous media}.
\newblock {\em Journal of Non-Newtonian Fluid Mechanics}, 235:76--82, 2016.

\bibitem{SalehiRS2019}
A.~Salehi-Shabestari; M. Raisee;~K. Sadeghy.
\newblock {Effect of a waxy crude oil's yield stress on the coning phenomenon:
  a numerical study}.
\newblock {\em Journal of Porous Media}, 22(1):21--35, 2019.

\bibitem{GargBSHJ2021}
A.~Garg; N. Bergemann; B. Smith; M. Heil;~A. Juel.
\newblock {Fluidisation of yield stress fluids under vibration}.
\newblock {\em Journal of Non-Newtonian Fluid Mechanics}, 294:104595, 2021.

\bibitem{PourzahediF2024}
A.~Pourzahedi;~I.A. Frigaard.
\newblock {A network model for gas invasion into porous media filled with
  yield-stress fluid}.
\newblock {\em Journal of Non-Newtonian Fluid Mechanics}, 323:105155, 2024.

\bibitem{SochiThesis2007}
T.~Sochi.
\newblock {\em {Pore-Scale Modeling of Non-Newtonian Flow in Porous Media}}.
\newblock PhD thesis, Imperial College London, 2007.

\bibitem{SochiB2008}
T.~Sochi;~M.J. Blunt.
\newblock {Pore-scale network modeling of Ellis and Herschel-Bulkley fluids}.
\newblock {\em Journal of Petroleum Science and Engineering}, 60(2):105--124,
  2008.

\bibitem{SochiYield2010}
T.~Sochi.
\newblock {Modelling the Flow of Yield-Stress Fluids in Porous Media}.
\newblock {\em Transport in Porous Media}, 85(2):489--503, 2010.

\bibitem{SochiElasticYield2013}
T.~Sochi.
\newblock {The Yield Condition in the Mobilization of Yield-Stress Materials in
  Distensible Tubes}.
\newblock {\em Central European Journal of Physics}, 12(8):532--540, 2014.

\bibitem{SochiYieldBal2013}
T.~Sochi.
\newblock {Yield and Solidification of Yield-Stress Materials in Rigid Networks
  and Porous Structures}.
\newblock 2015.
\newblock arXiv:1311.2644.

\bibitem{SochiStress1D2015}
T.~Sochi.
\newblock {Using the stress function in the flow of generalized Newtonian
  fluids through pipes and slits}.
\newblock 2015.
\newblock arXiv:1503.07600.

\bibitem{SochiStressFunc22015}
T.~Sochi.
\newblock {Using the stress function in the flow of generalized Newtonian
  fluids through conduits with non-circular or multiply connected cross
  sections}.
\newblock 2015.
\newblock arXiv:1509.01648.

\bibitem{BarnesW1985}
H.A. Barnes;~K. Walters.
\newblock {The yield stress myth?}
\newblock {\em Rheologica Acta}, 24(4):323--326, 1985.

\bibitem{Astarita1990}
G.~Astarita.
\newblock {The engineering reality of the yield stress}.
\newblock {\em Journal of Rheology}, 34(2):275--277, 1990.

\bibitem{Evans1992}
I.D. Evans.
\newblock {On the nature of the yield stress}.
\newblock {\em Journal of Rheology}, 36:1313--1316, 1992.

\bibitem{Barnes1999}
H.A. Barnes.
\newblock {The yield stress---a review or `$\pi\alpha\nu\tau\alpha \
  \rho\varepsilon\iota$'---everything flows?}
\newblock {\em Journal of Non-Newtonian Fluid Mechanics}, 81(1):133--178, 1999.

\bibitem{MollerMB2006}
P.C.F. M{\o}ller; J. Mewis;~D. Bonn.
\newblock {Yield stress and thixotropy: on the difficulty of measuring yield
  stresses in practice}.
\newblock {\em Soft Matter}, 2(4):274--283, 2006.

\bibitem{Renardy2010}
M.~Renardy.
\newblock {The mathematics of myth: Yield stress behavior as a limit of
  non-monotone constitutive theories}.
\newblock {\em Journal of Non-Newtonian Fluid Mechanics}, 165(9-10):519--526,
  2010.

\end{thebibliography}
\phantomsection \addcontentsline{toc}{section}{Nomenclature}

\section*{\label{Nomenclature}Nomenclature}

\noindent %
\begin{longtable}[l]{ll}
2D, 3D & two dimensional, three dimensional\tabularnewline
\addlinespace[0.05cm]
$a$ & length of side of equilateral triangle (m)\tabularnewline
\addlinespace[0.05cm]
$a,b$ & semi-major and semi-minor axes of ellipse (m)\tabularnewline
\addlinespace[0.05cm]
$a,b$ & semi-length and semi-width of rectangle (m)\tabularnewline
\addlinespace[0.05cm]
$a,b$ & outer and inner radii of circular annulus (m)\tabularnewline
\addlinespace[0.05cm]
$p$ & pressure (Pa)\tabularnewline
\addlinespace[0.05cm]
$r$ & radius (m)\tabularnewline
\addlinespace[0.05cm]
$x,y,z$ & coordinate variables (usually spatial coordinates)\tabularnewline
\addlinespace[0.05cm]
$\tau_{xz},\tau_{yz},\tau_{rz},\tau_{\theta z}$ & shear stress components (Pa)\tabularnewline
\addlinespace[0.05cm]
\end{longtable}
\end{document}